\documentclass{article}[14pt,a4paper]
\usepackage{hyperref}

\usepackage{amsthm,amsmath,amssymb}
\usepackage[left=2.5cm, right=3cm, top=2cm]{geometry}
\usepackage{graphicx}
\usepackage{url}
\usepackage{caption}
\usepackage{graphicx}		
\usepackage{subcaption}

\title{Minimum error correction-based haplotype assembly: considerations for long read data} 

\author{Sina Majidian$^1$, Mohammad Hossein Kahaei$^{1*}$ and Dick de Ridder$^2$ }
\date{\small $^1$School of Electrical Engineering, Iran University of Science \& Technology, Narmak, 16846-13114, Tehran, Iran.\\
$^2$Bioinformatics Group, Wageningen University, Droevendaalsesteeg 1, 6708PB, Wageningen, The Netherlands.\\
$^*$ Corresponding author.}

\begin{document}
\maketitle

\begin{abstract}
The single nucleotide polymorphism (SNP) is the most widely studied type of genetic variation. A haplotype is defined as the sequence of alleles at SNP sites on each haploid chromosome. Haplotype information is essential in unravelling the genome-phenotype association. Haplotype assembly is a well-known approach for reconstructing haplotypes, exploiting reads generated by DNA sequencing devices. The Minimum Error Correction (MEC) metric is often used for reconstruction of haplotypes from reads. However, problems with the MEC metric have been reported. Here, we investigate the MEC approach to demonstrate that it may result in incorrectly reconstructed haplotypes for devices that produce error-prone long reads. Specifically, we evaluate this approach for devices developed by Illumina, Pacific BioSciences and Oxford Nanopore Technologies.  We show that imprecise haplotypes may be reconstructed with a lower MEC than that of the exact haplotype. The performance of MEC is explored for different coverage levels and error rates of data.  Our simulation results reveal that in order to avoid incorrect MEC-based haplotypes, a coverage of 25 is needed for reads generated by Pacific BioSciences RS systems. 
\end{abstract}


\section{Introduction}
Among the various types of genetic variations, single nucleotide polymorphisms (SNPs) are the most widely studied among others in genome wide association studies (GWAS). The genome of diploids like humans consists of two homologous pairs: the paternal and maternal chromosomes. A haplotype, the sequence of alleles at SNP sites on each homologous chromosome, can be measured through direct experiments or can be reconstructed by computational approaches \cite{sny,schwartz}. Due to the high cost of experimental methods, the computational techniques have attracted more attention. These techniques can be categorized as phasing or assembly approaches. Phasing makes use of the genotypes of multiple individuals to infer the haplotype. In the haplotype assembly approach, sets of reads generated by DNA sequencing devices are exploited for haplotype reconstruction. While haplotype assembly can be performed for a single individual, phasing cannot. Moreover, phasing is difficult in the presence of low-frequency and \emph{de novo} variants.

The history of DNA sequencing technologies consists of three generations. Firstly, the low-throughput Sanger sequencing machines were built in the late 1980s, thanks to the invention of the chain termination procedure. Subsequently, multiplexing strategies were used for the development of the so-called second generation technologies of the early 2000s.  Today, Illumina is the dominant platform of this second generation, providing massively high throughput, up to billions of reads, with a length of a few hundred bases and an error probability lower than 0.001 \cite{shen}. Utilizing such short reads incurs limitations, precluding assembly of repetitive regions and detection of structural variants larger than read length. The third-generation of sequencing technology, namely single-molecule sequencing as provided by Pacific Biosciences (PacBio) and Oxford Nanopore Technologies (ONT), produces exceptionally long reads of up to a million bases.  The bottleneck of this third-generation technology is the low per-base accuracy in comparison to that of the second generation, such that the error probability may exceed 0.1 \cite{goodwin}. Both  second and third generation  sequencing technologies have been used for haplotype assembly. 

Although the sequencing reads provided by all above-mentioned technologies do not keep track of the haplotypic origin of reads, a haplotype assembly algorithm tends to reconstruct the haplotypes using overlaps among reads. In the absence of sequencing errors, this is a trivial problem to solve. A simple bipartitioning scheme can be used to divide reads into two groups corresponding to two haplotypes, such that those reads in each group do not conflict. But in real cases, the presence of errors makes the problem computationally hard to solve. Several criteria have been proposed in the literature, including minimum fragment removal (MFR), minimum SNP removal (MSR) and minimum error correction (MEC) \cite{lancia}. The idea behind MFR is to find the minimum number of reads containing errors, which should then be removed. The heuristic algorithms for solving this model are time-consuming and not suitable for low coverage input data. 
In MSR-based algorithms, several SNP positions are removed to make haplotyping possible. Thus, the haplotypes contain some gaps, leading to a high SNP missing rate, which is undesired.

The dominant objective function utilized for the haplotype assembly problem is the MEC, also known as the minimum letter flip \cite{schwartz}. This function is also used in evaluating the performance of different haplotype reconstruction algorithms \cite{haptree,bansal}. Minimizing the MEC function can be rewritten as a MAXCUT problem, which is NP-hard, leading to a large number of heuristic algorithms \cite{lancia}. Some examples include the HapCUT algorithm (which iteratively computes max-cuts of a read graph \cite{hapcut}), a branch-and-bound genetic algorithm approach \cite{wang05}, an integer linear programming approach \cite{chen} and a clustering approach \cite{das}, as well as multiple dynamic programming approaches \cite{kuleshov,he,deng,bonizzoni}.

Despite the existence of all these methods utilizing the MEC for haplotype reconstruction, it is crucial to note that this criterion may fail to identify the exact haplotype when there is a high error rate in the reads \cite{wang05,zhang}. In addition, a negative correlation between the haplotype accuracy and the MEC has already been reported in \cite{refhap}, as discussed in the Results section.
While this issue has been mentioned briefly in previous studies, it has never been systematically investigated in an effort to understand the implications across different sequencing platforms.

In this work, we provide insight into the MEC function to clarify the above ambiguities. The following section presents the fragment matrix model, defines MEC and introduces two theorems regarding MEC performance. The performance curve for MEC is introduced and discussed in the Results section. Furthermore, several DNA sequencing devices are evaluated based on their characteristics, including the error probability values. Finally, simulations of long and short reads are provided to explore practical consequences.

\section{Methods}
For diploids, haplotype assembly is the process of reconstructing two haplotypes from overlapped aligned reads. Throughout this paper, we only consider bi-allelic SNPs - that is, SNPs with only one alternative allele against the reference allele \cite{hapcut,kuleshov}.
Below we describe the construction of the fragment matrix. Prior to this construction, we remove those reads that cover less than two SNP sites, because these are not informative for haplotype assembly. Non-SNP bases of each read are also omitted. 

\subsection{Fragment matrix model}  
We  assume that there are $N$ reads obtained from both chromosomes. For a haplotype with the length of $l$, an $N\times l$ fragment matrix $\boldsymbol R$ is constructed whose rows embed the reads and whose columns correspond to the heterozygous SNP sites \cite{si,maj}. The SNP sites not covered by the reads are coded with zero.  Then, bases of reads are converted to $-1$ (alternative allele) or $1$ (reference allele), assuming bi-allelic SNPs. 

As an example of an error-free case, consider the first exon of HLA-A , a gene on chromosome 6 -with NCBI reference sequence number NG\_029217.2. Its first 40 bases are presented in Fig.\@1a. It contains five bi-allelic  SNP sites (refSNP):  C/T (rs753601428), C/G  (rs529070997), G/T (rs41560714), A/C (rs551138783) and A/G (rs778615037). The procedure of constructing the fragment matrix is depicted in Fig.\@1d. In this example, the exact haplotypes that should be reconstructed by the haplotype assembly algorithms are \{CGTAG\} and \{TCGCA\}.

\quad

\begin{figure}[!ht]
\includegraphics[width=\textwidth]{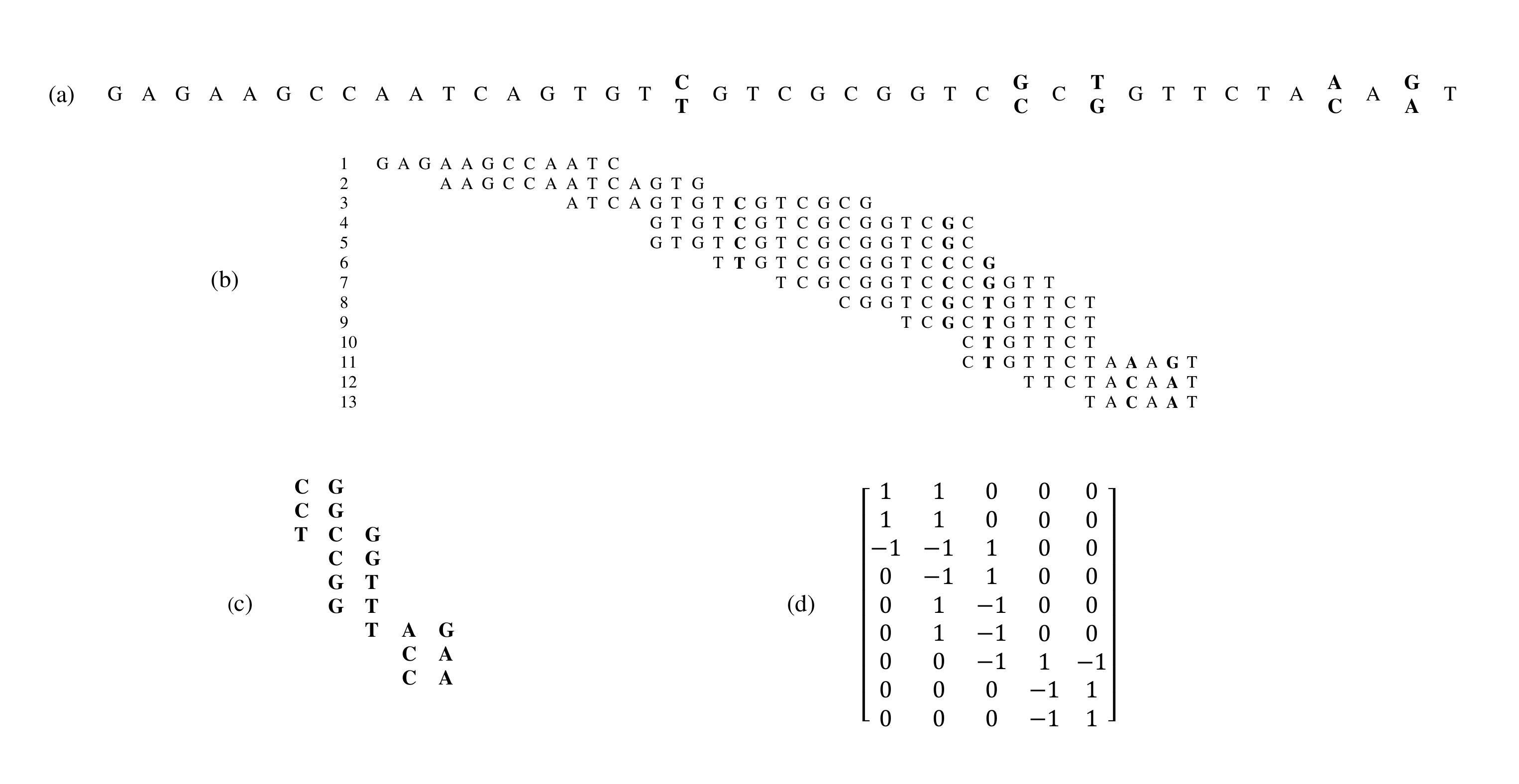}
\caption{An example of fragment matrix model for the first 40 bases of exon 1 of HLA-A gene. This gene is located on chromosome 6 with NCBI reference sequence number  NG\_029217.2. It  contains 5 bi-allelic SNP sites (refSNP): C/T (rs753601428), C/G  (rs529070997), G/T (rs41560714), A/C (rs551138783) and A/G (rs778615037). a) An example of homologous chromosomes in which the SNP sites are indicated in bold, b) an example of aligned reads,  c) the fragments after removing non-informative reads and non-SNP bases and d) the constructed fragment matrix.}
\end{figure}

\newpage

The fragment matrix $\boldsymbol R$ can be modeled using a matrix completion approach \cite{si,maj}.  In the error-free case, $\boldsymbol R$ is a partially observed matrix modelled as
\begin{equation}
\boldsymbol R=P_\Omega(\boldsymbol M), 
\end{equation}
where $\boldsymbol M$ is the completed version of matrix $\boldsymbol R$ (see Appendix B for more details). $P_\Omega$ is the observation operator defined as
\begin{equation}
[P_\Omega(\boldsymbol M)]_{ij}=
\left\{\begin{array}{l}
\boldsymbol M_{ij}, \quad (i,j)\in \Omega,\\
0, \quad \text{\quad  \; otherwise,} \end{array}\right.	
\end{equation}
in which $\Omega$ is the set of indices of known entries. In order to generalize the model to the more realistic case allowing erroneous entries, we use an additive measurement error model  inspired by \cite{das,si,maj}:
\begin{equation}
\boldsymbol R=P_\Omega(\boldsymbol M)+\boldsymbol E.
\end{equation}

To define the error matrix $\boldsymbol E$, we should first clarify what we mean by an error. A substitution error is the conversion of a DNA base to one of the other three possible bases during the sequencing procedure. 
As mentioned earlier, during fragment matrix construction, only two bases (reference and alternative alleles) for each SNP site are permitted and other possible bases are ignored; as a result, a substitution to the ignored bases does not affect the entries of the fragment matrix. Accordingly, we introduce the term bi-allelic substitution, or simply bi-substitution to make it distinguishable from generally defined substitution. A bi-substitution error occurs when a reference allele is converted to the alternative allele or vice versa. Consequently, an error in the entries of $P_\Omega(\boldsymbol M)$ is simplified as a change from $-1$ to $1$ or vice versa. This can be formulated as an addition of $2$ (or $-2$) to each erroneous entry of $P_\Omega(\boldsymbol M)$ which is represented in error matrix $\boldsymbol E$.
We assumed that each non-zero entry of $\boldsymbol R$ is erroneous with a probability of $p_e$, the bi-substitution error probability, independent of the other entries. This value equals one third of the substitution error probability of the sequencing device $p_s$.

\subsection{MEC definition}
If the reads contain no errors, the corresponding rows of fragment matrix are compatible with each other and haplotypes are extracted using a simple clustering technique. 
However, in practice, sequencing devices may produce erroneous reads due to which the compatibility of reads  is lost. To cope with this problem, the MEC approach is employed by inverting the sign of some entries of  the fragment matrix to make it compatible \cite{wang05}:

\begin{enumerate}
\item Find the minimum number of entries of $\boldsymbol R$ that should be inverted to make the fragment matrix compatible.
\item Cluster the rows of the augmented fragment matrix and reconstruct the haplotype.
\end{enumerate}

For the fragment matrix $\boldsymbol R$ with the dimension of $N \times l$ and the candidate $1\times l$ vector $\boldsymbol h_c$ as the haplotype, the MEC function is calculated as 
\begin{equation}
\text{MEC}(\boldsymbol R,\boldsymbol  h_c)= \sum_{i=1}^{N} \text{min}\{D(\boldsymbol r_i,\boldsymbol h_c),D(\boldsymbol r_i,-\boldsymbol h_c)\},
\end{equation}
in which $\boldsymbol r_i$ is the $i^{th}$ row of $\boldsymbol R$ and the  extended Hamming distance (EHD) is defined as $D(\boldsymbol a,\boldsymbol b) =\sum_{k=1}^{l} d(\boldsymbol a(k),\boldsymbol b(k))$ \cite{chen,hapcut}. Furthermore, $d(\cdot,\cdot)$ is a mismatch indicator which penalizes its dissimilar arguments by one:
\begin{equation}
d(a,b)=  \left\{\begin{array}{l}
1, \text{ if } a\neq0\text{ \& }b\neq0\text{ \& }a\neq b\\
0, \text{ otherwise.} \end{array}\right.
\end{equation}
Therefore, the EHD function represents the number of mismatches between two vectors. From this point of view, $\text{MEC}(\boldsymbol R,\boldsymbol  h_c)$ indicates the whole number of mismatches between each row of $\boldsymbol R$ and the vector $\boldsymbol  h_c$.
 It is notable that  the function  $D (\cdot,\cdot)$ is not a distance from the  mathematical point of view, though it is named as such (See Appendix A).

\subsection{Analysis of MEC performance}
Consider $\boldsymbol h_{opt}$ as an optimal solution resulting from a given method by minimizing the MEC function. The question arises: does minimizing this function guarantee reaching the exact haplotypes (i.e., the true haplotypes of the individual)? In Theorem 1, we demonstrate not only that this solution offers no guarantee of finding the exact haplotype, but also that the MEC function will not lead to the exact haplotype. 

\quad 

\noindent \textbf{Theorem 1.} There exists a vector $\boldsymbol h_d$ different from the exact haplotype $\boldsymbol h_{ex}$ with a lower MEC, when the $k^{th}$ column of the fragment matrix, $\boldsymbol R$, contains some erroneous entries whose number $E^{(k)}$ is greater than half of its coverage. In a mathematical expression:
\begin{equation}
\text{If }\exists k:  \frac{E^{(k)}}{c^{(k)}}> \frac{1}{2}  \quad \text{then} \quad \exists \boldsymbol h_d \neq \boldsymbol h_{ex} : \text{MEC}(\boldsymbol R,\boldsymbol h_d)<\text{MEC}(\boldsymbol R,\boldsymbol h_{ex}),
\end{equation}
where $c^{(k)}$  is the coverage (or the read depth) of the $k^{th}$ SNP site. The coverage indicates the number of reads that covers the SNP and is equal to the number of known entries of the $k^{th}$ column of $\boldsymbol R$.  We conclude that the ratio $E^{(k)}/ c^{(k)}$, called the bi-substitution rate, plays a key role in the evaluation of a sequencing device. From a practical perspective, $E^{(k)}$, the number of nonzero values of the $k^{th}$ column of $\boldsymbol{E}$, represents the number of bi-substitutions at the corresponding genomic position (see section Fragment matrix model). 
The proof of Theorem 1 is presented in Appendix B. The core idea of the proof is to consider $\boldsymbol h_{d}$ equal to $\boldsymbol h_{ex}$ except in its $k^{th}$ entry, whose sign is inverted. This guarantees a lower MEC. 

Note that if the antecedent is not satisfied, the MEC approach works properly. In practice, fulfilling the antecedent of Theorem 1 is a major point to be investigated further. To explore this point, Theorem 2 presents the probability of the antecedent not occurring.  

\quad

\noindent \textbf{Theorem 2.}  The probability of obtaining a minimum MEC value for the exactly correct haplotype  $(P\{\text{c-MEC}\})$ is equal to
\begin{equation}
P\{\text{c-MEC}\}=\prod_{j=1}^{l}\left\{\sum_{n=0}^{\lfloor \frac{c^{(j)}}{2}\rfloor}{ {c^{(j)}\choose {n}} p_e^n (1-p_e)^{c^{(j)}-n}} \right\}.
\end{equation}
in which $p_e$ is the bi-substitution error probability.

\quad 

\noindent Proof:
According to Theorem 1, the MEC approach works properly when the number of erroneous entries of each column is lower than half of its corresponding coverage. Based on the above assumption, the number of erroneous entries of each column of $\boldsymbol R$ is independent of the other columns. Then, we have:
\begin{equation}
P\{\text{c-MEC}\}=\prod_{j=1}^{l}{ P\left\{ E^{(j)}<\frac{c^{(j)}}{2} \right\} }.
\end{equation}
An erroneous entry gets the opposite sign  due to the bi-allelic assumption. This follows a Bernoulli distribution of $\pm1$ with the probability of error $p_e$. Thus, the number of errors in the $j^{th}$ column  follows a Binomial distribution given by $P\left\{ E^{(j)}=k\right\}={c^{(j)}\choose {k}} p_e^k (1-p_e)^{c^{(j)}-k}$. Therefore, we can write:
\begin{equation}
P\left\{ E^{(j)}<\frac{c^{(j)}}{2}\right\}= \sum_{k=0}^{\lfloor c^{(j)}/2\rfloor}{ {c^{(j)}\choose {k}} p_e^k (1-p_e)^{c^{(j)}-k}}.
\end{equation}
Accordingly, using (8) and (9) the proof of Theorem 2 is complete.

\section{Results}
\subsection{Performance curves of MEC}
 The outcome of Theorem 2 is calculated for various scenarios with different probabilities of error and coverage levels. This is done by introducing performance curves for MEC. The $y$-axis indicates the probability of obtaining a correct MEC $P\{\text{c-MEC}\}$ and the $x$-axis the bi-substitution error probability $p_e$. 

In practice, the average coverage of input data provided for haplotype assembly varies from very low to very high levels. Based on the the existing literature on coverage distribution among different genomic positions \cite{lander, klambauer,si}, we consider two different distributions, including Poisson and quasi-uniform (i.e., the analogue of the uniform distribution defined for a discrete random variable), as well as constant coverage levels. The error probability of various datasets may also differ dramatically due to the specifications of the DNA sequencer.  

In Fig.\@2a, the performance curve, $P\{\text{c-MEC}\}$ versus $p_e=[0.0001,0.5]$ is presented for different coverage values. In three cases, we consider $c^{(j)}=2$, $10$ and $100$ for $j=1,\ldots, l$, respectively. Next, $c^{(j)}$s are defined randomly by the quasi-uniform distribution over three different intervals $[1,2]$, $[1,10]$ and $[1,100]$. In addition,  MEC performance is investigated for coverage values of SNP sites with the Poisson distribution with mean $\lambda=2$, $10$ and $100$.
Furthermore, Fig.\@2b  displays  $P\{\text{c-MEC}\}$ for different lengths of haplotypes $l=\{100,10k,1M\}$ and coverage values $c=\{2,10,30\}$.

\begin{figure}[!ht]
\begin{subfigure}{0.95\textwidth}
\includegraphics[width=.95\textwidth]{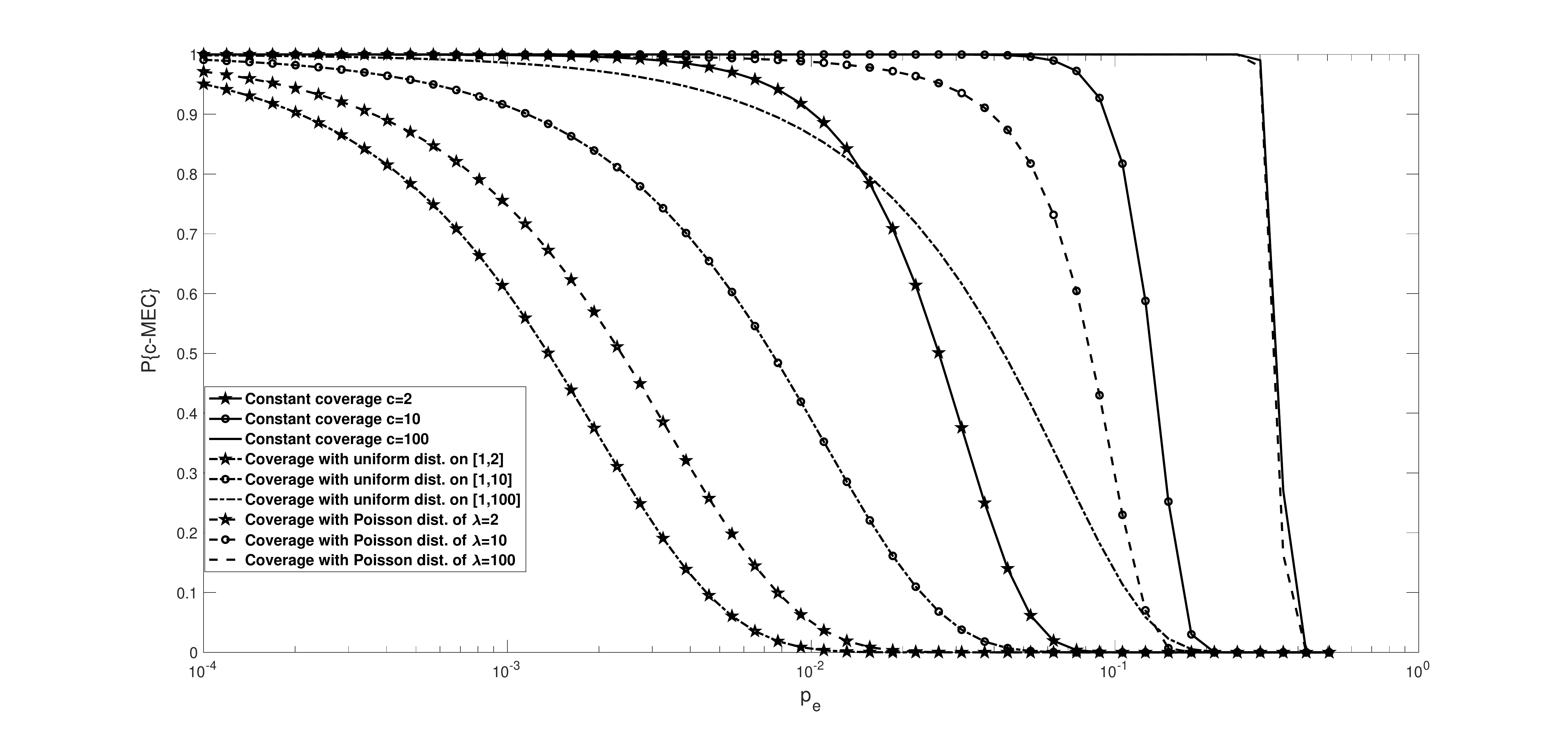}
\caption{ }
\end{subfigure}
\begin{subfigure}{0.95\textwidth}
\includegraphics[width=.95\textwidth]{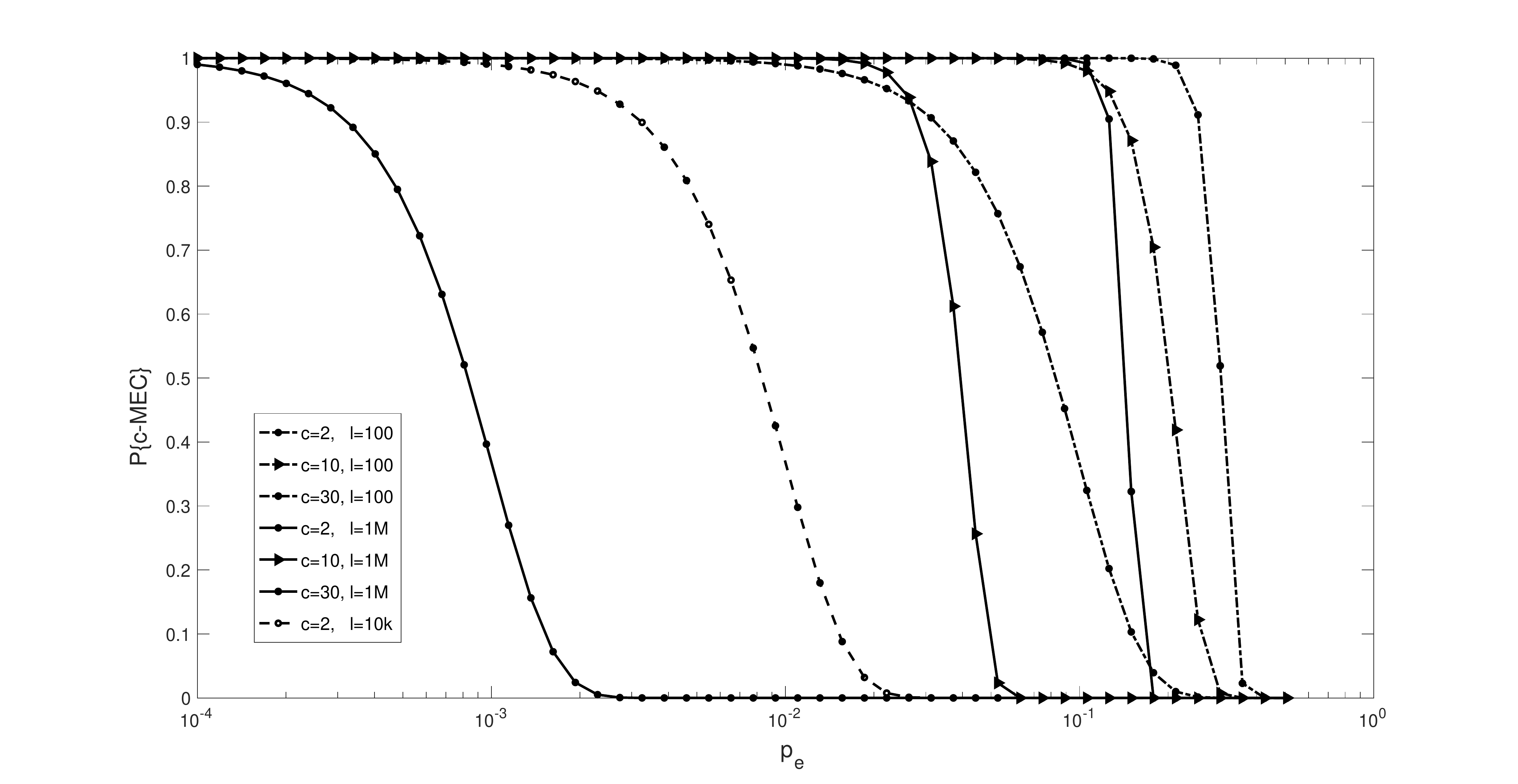}
\caption{ }
\end{subfigure}
\caption{Performance curves of MEC approach:  a) Comparison of $P\{\text{c-MEC}\}$ for different coverage levels (constant $c=\{2,10,100\}$, quasi-uniform over $ c=\{[1,2],[1,10],[1,100]\}$ and Poisson distribution with mean $\lambda=\{2,10,100\}$),  b) Comparison of $P\{\text{c-MEC}\}$ for different haplotype lengths $l=\{100,10k,1M\}$ and different coverage values $c=\{2,10,30\}$.}
\end{figure}

\newpage
In Fig.\@2a, it is seen that $P\{\text{c-MEC}\}$ is inversely proportional to the sequencing error probability $p_e$. Additionally, depending on the coverage distribution, each $P\{\text{c-MEC}\}$ begins to drop after a particular threshold. For example, for the Poisson distribution with mean $\lambda=10$ and $l=1k$, this threshold is $p_e=2\%$.  In this case, the MEC approach is unable to reconstruct the exact haplotype for $p_e>2\%$. This problem arises when the number of errors in column is more than half of its coverage, as expressed in Theorem 1. The existence of such a column is more likely as the error probability increases.
Fig.\@2b presents our investigation on the effect of the haplotype length on $P\{\text{c-MEC}\}$. It demonstrates that a higher haplotype length $l$ leads to incorrect haplotypes at a lower bi-substitution error probability $p_e$.


\subsection{Evaluation of sequencing technologies: theory}
Here, we analyze the MEC for different DNA sequencing devices based on our reasoning. 
Table 1 presents the results of the evaluation of different devices launched by Illumina, PacBio and ONT. 
For each device, the evaluation employs the typical number of reads per run, the read length and  error probability as reported in literature \cite{goodwin,tyl,pb}.
In order to provide a fair comparison, we set the coverage value at 10. To calculate the number of runs needed (denoted by $n$) for such coverage, we used the averaged coverage formula, the Lander-Waterman equation, as following: 
\begin{equation}
c_{a}= \frac{l_r N_t}{G},
\end{equation}
where $l_r$, $N_t$ and $G$ show the read length, the total number of reads per run and the human genome length, respectively.



The applicability of the MEC approach for data generated by each device is reported in the last column of Table 1, based on the value of $P\{\text{c-MEC}\}$. This shows that the MEC criterion works well for short reads produced by Illumina devices, but not for long reads produced by PacBio or ONT.  A larger value of $n$ corresponds to a higher sequencing cost for each device.  It should be noted that for each run, long-read devices are far more expensive than short-read devices. 


\begin{table}[!htp]
\center
\caption{ Comparison MEC applicability of different sequencing devices for the substitution error probability $p_s$, the total number of reads $N_t$ in millions, the read length $l_r$ and the number of runs $n$ needed for a coverage of $10$. For Illumina technology, the read lengths corresponds to the paired-end setting.}  
\begin{tabular}{|l|c|c|c|c|c|c|} \hline
Device&   $p_s$ & $N_t$ & $l_r$  & $n$ & $P\{\text{c-MEC}\}$  & MEC applicability  \\ \hline
Illumina MiSeq V3         &0.001 &50    &300    &2  &0.97 &Yes \\\hline
Illumina HiSeq 4000       &0.001 &2500  &150    &1  &0.97 &Yes \\\hline
Illumina HiSeq X          &0.001 &2600  &150    &1  &0.97 &Yes\\\hline
Pacific BioSciences RS II &0.06   &0.055 &20k    &30 &0.23 &No \\\hline
Pacific BioSciences Sequel&0.06   &0.35  &12k    &10 &0.23 &No \\\hline
Oxford Nanopore MinION    &0.02   &0.1   &200k   &2  &0.42 &No \\\hline
\end{tabular}
\end{table}

\subsection{Evaluation of sequencing technologies: simulations}
We run various simulations to provide a deeper understanding of MEC-based haplotype assembly.
First, using DNA sequencing data, we estimate how often MEC failures can occur based on Theorem 1.
The accuracy of the reconstructed haplotype is also investigated in terms of switch error rate and haplotype block length. 

\subsubsection{On the satisfaction of Theorem 1}
Here, we inspect the effect of short and long sequencing reads along with their corresponding error profiles for the satisfaction of antecedent of Theorem 1. To do so, we use  the bi-substitution rate defined in the Methods section.

We briefly present the details of our simulations. We consider the 21st chromosome of the human genome (GRCh38) \cite{grch} as the reference DNA sequence. Bi-allelic SNPs are introduced at a rate of one in a thousand bases \cite{levy} across the mentioned reference using haplo-generator, part of the haplosim package \cite{mot}. For generating PacBio long reads, we use the PBSIM package \cite{pb} in which the PacBio error profile is used. Then, we align the reads using minimap2 \cite{mini}.  We run the ART pacakge \cite{art} for generating short paired-end reads and Burrows-Wheeler Aligner (BWA) \cite{bwa} for aligning them. We sort the aligned reads using the samtools package \cite{sam}.  Afterwards, using the mpileup subprgram of samtools \cite{sam}, alleles for each position are extracted from the sorted aligned reads. Then, the required statistics for all introduced SNPs are calculated.  

For both Illumina reads and PacBio long reads, the number of SNPs with a bi-substitution rate of greater than or equal to $0.5$ are depicted in Fig.\@3. 
In Appendix B, we depict the histogram of bi-substitution rates of SNP sites.  For coverage values up to 25 for PacBio data, there are some positions in which the bi-substitution rate is greater than $0.5$. This leads to the satisfaction of the antecedent of Theorem 1 and thus MEC failure. When we set the coverage greater than or equal to $c=30$, no SNP site with high bi-substitution rate remains.

\begin{figure}[!ht]
\centering
\includegraphics[width=.6\textwidth]{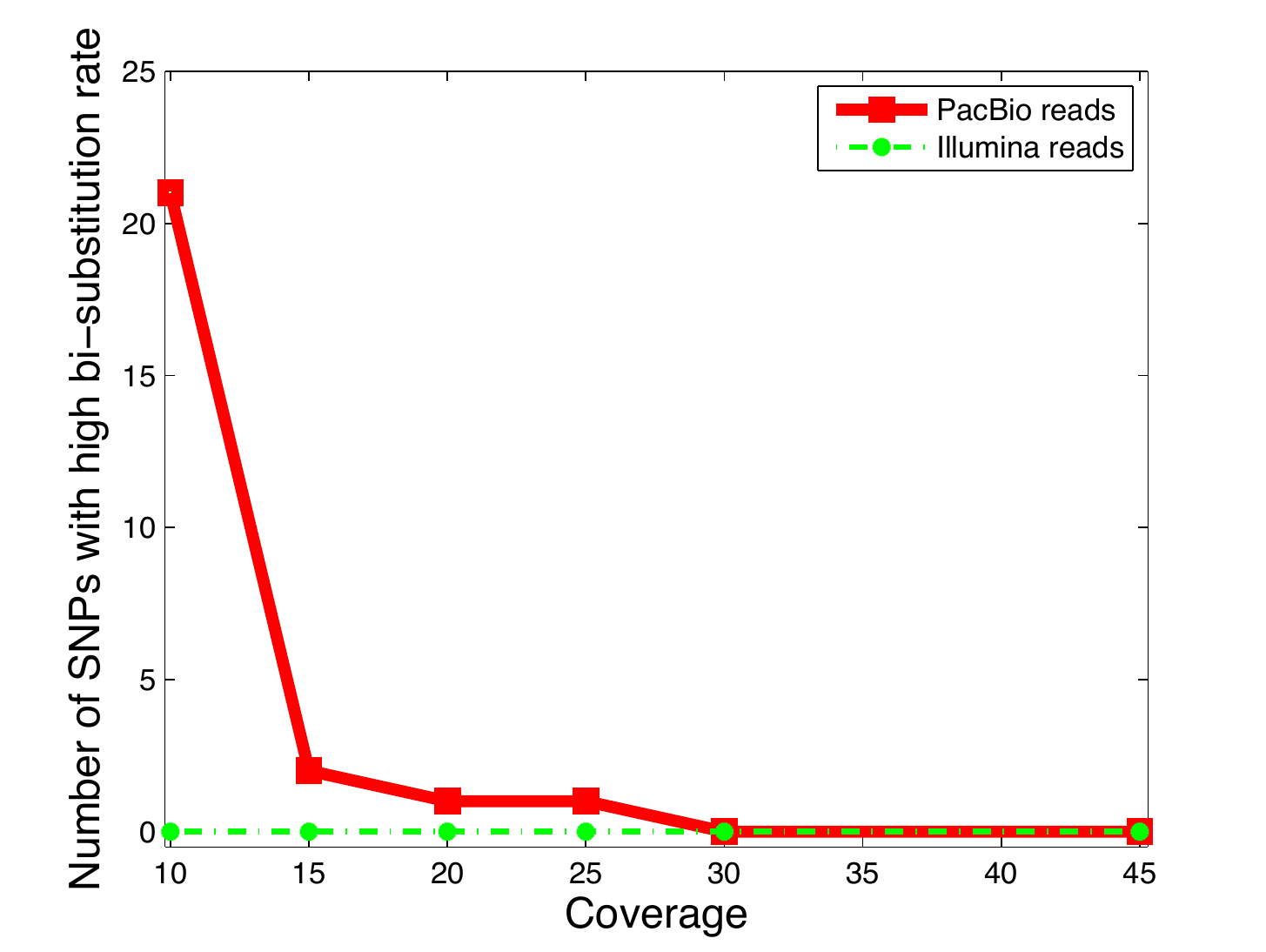}
\caption{Number of SNPs with bi-substitution rate of greater than or equal to $0.5$  (high bi-substitution) for Illumina reads and PacBio long reads at different coverage levels.}
\end{figure}

\subsubsection{Haplotype reconstruction accuracy}
We now examine the direct effect of coverage on the accuracy of the reconstructed haplotype. We utilize the well-known HapCUT algorithm as a MEC-based haplotype assembly method.

The output of HapCUT consists of haplotype blocks, whose continuity can be evaluated by calculating the average block length. Larger haplotype blocks, indicating that haplotypes are reconstructed more continuously, are of interest.
To evaluate the accuracy of the reconstructed haplotype, we calculate the switch error rate by dividing the number of switch errors by the haplotype length. A change in the parental origin of an allele compared to the previous allele is called a switch error.

The switch error rate and average block length of the haplotype reconstructed by HapCUT  are depicted for different coverage values from $c=10$ to $45$ in Fig.\@4a and b, respectively.  The results are provided for 20 independent generated datasets.
As seen in both figures, by increasing the coverage, the accuracy and continuity of the reconstructed haplotype increases. For a dataset with low coverage, specifically lower than 25 per haploid, not only are there many switches but the reconstructed haplotype is highly fragmented as well. This corroborate the findings in Fig.\@3.






\begin{figure}[!ht]
\centering
\begin{subfigure}[b]{0.49\textwidth}
\includegraphics[width=\textwidth]{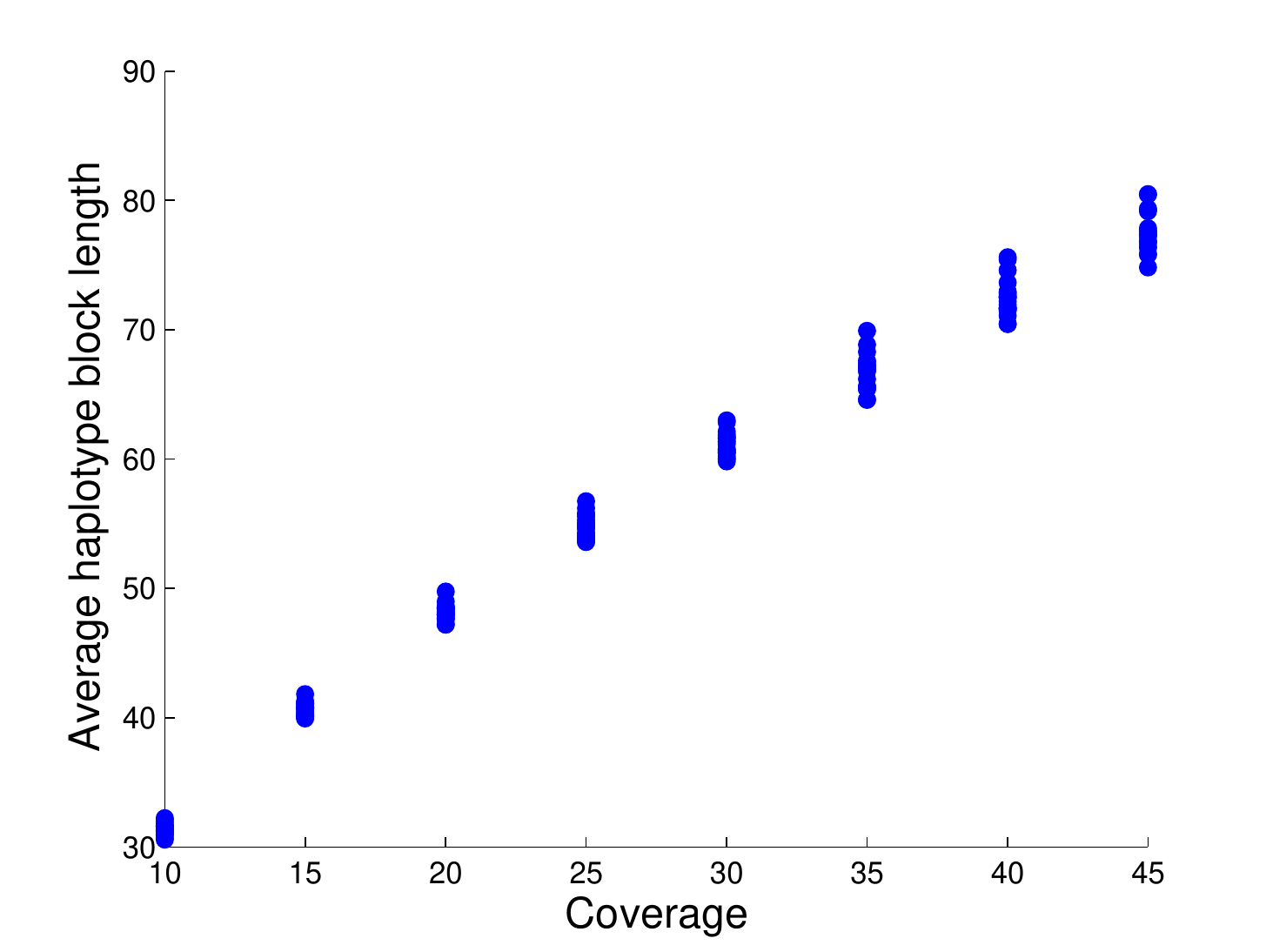}
\caption{ }
\end{subfigure}
\begin{subfigure}[b]{0.49\textwidth}
\includegraphics[width=\textwidth]{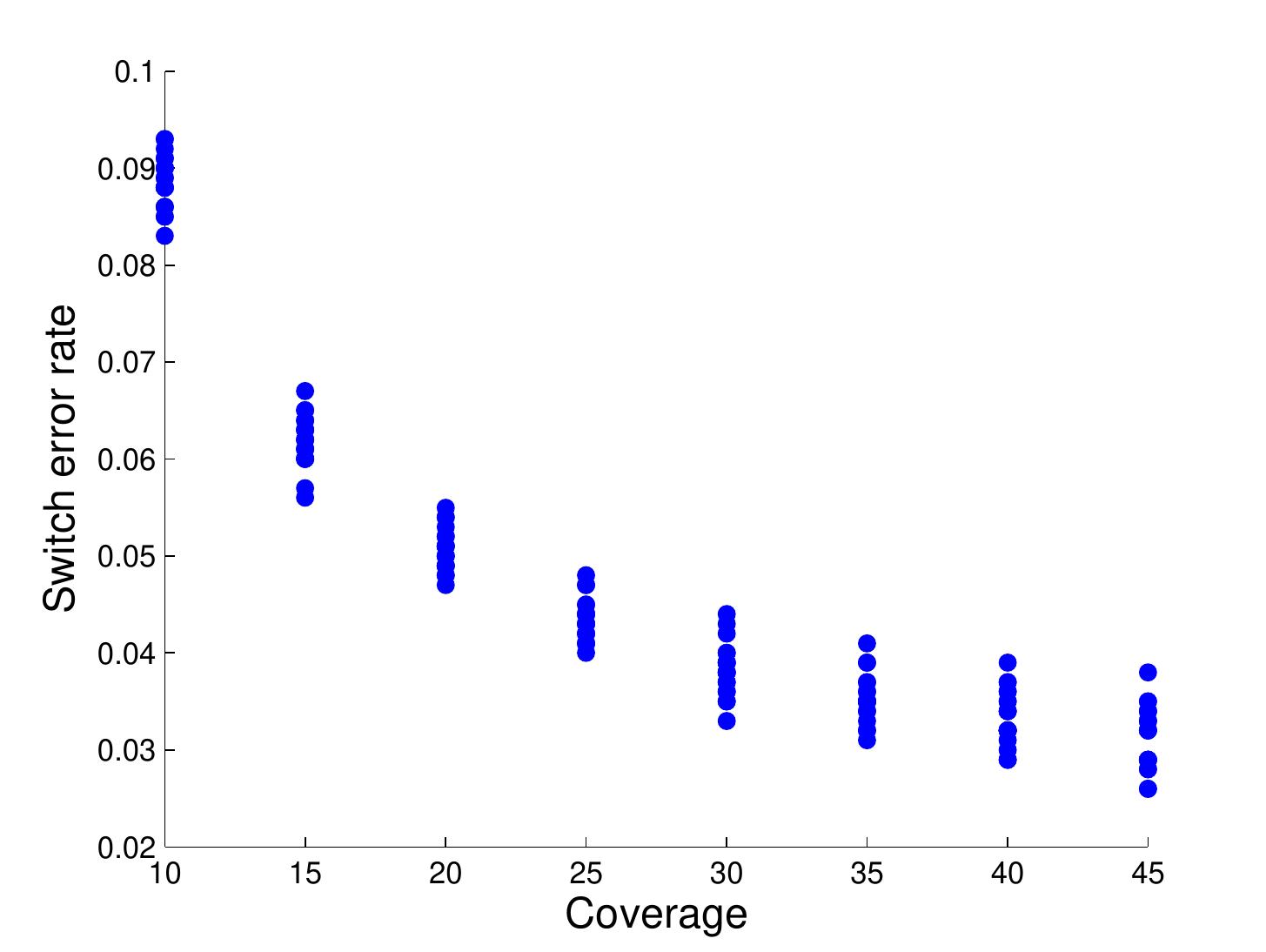}
\caption{ }
\end{subfigure}
\caption{Accuracy of reconstructed haplotypes using HapCUT in terms of average haplotype block length and  switch error rate.}
\end{figure}

\newpage

\section{Discussion and conclusion}
The issue addressed in this paper has been recognized previously by Duitama \emph{et al.} \cite{refhap}, who note that a candidate haplotype with lower MEC is associated with lower reconstruction accuracy. This result can be predicted from the model we described.
We investigated the reliability of the MEC approach for haplotype assembly. We demonstrate that in some practical circumstances, an imprecise haplotype may be reconstructed with a lower MEC than that of the exact haplotype. The theoretical MEC performance curves were  obtained for different coverage values and error rates. Based on our analyses, we evaluated some DNA sequencing devices by the MEC criterion. It was found that this approach can generate misleading results for low-coverage error-prone long reads generated by Pacific BioSciences and Oxford Nanopore Technologies platforms. In order to address this issue, one should exploit a high coverage for long reads.  The results provided in this study suggest that using MEC-based haplotype assembly methods on available long reads, reconstruction of the true haplotypes is not feasible for coverage lower than 25 per haploid (i.e., 50 overall). 

It should be noted that, while we assume errors to be an independent and identically distributed (iid), in reality this may not hold true, although this assumption has been used before widely \cite{haptree,trip}.  Though PacBio reads have no systematic error, errors in alignment and variant calling may exist due to high numbers of insertions and deletions. Acquiring comprehensive error models for all sequencing technologies is a difficult task and exploiting them in our model would make the derivation unfeasible. Therefore, we used an approach that is simplified yet close to reality. 

An important future direction for this work is to do a thorough research on the extent of the issues with MEC for the polyploid genome. In Appendix E, we present the MEC formula for polyploids and we show that MEC failure may also happen in a specific polyploid case.

\section*{Availability of data and materials}

The bash scripts and python codes are available at  \url{https://github.com/smajidian/MEC}. The reference human genome are downloaded from  \url{http://hgdownload.cse.ucsc.edu/goldenPath/hg38/chromosomes/chr21.fa.gz}. 

\section*{Funding}
The authors received no specific funding for this work.
\section*{Competing interests}
The authors have declared that no competing interests exist.

\section*{Author Contributions} 
Formal analysis: Sina Majidian. \\
Methodology: Sina Majidian, Mohammad Hossein Kahaei and Dick de Ridder\\
Supervision: Mohammad Hossein Kahaei and Dick de Ridder\\
Writing – original draft: Sina Majidian.\\
Writing – review \& editing: Mohammad Hossein Kahaei and Dick de Ridder



\section*{Appendices}
\renewcommand{\thesection}{\Alph{section}}
\setcounter{section}{0}
\numberwithin{equation}{section}

\section{Extended hamming distance}
The EHD function $D(\cdot,\cdot)$  is defined as  \begin{equation}
\text{D}: \{0,1,-1\}^l \times \{0,1,-1\}^l \rightarrow \mathbb{R}^+\cup \{0\},D(\boldsymbol a,\boldsymbol b) =\sum_{j=1}^{l} d(\boldsymbol a(j),\boldsymbol b(j)),
\end{equation}
where
\begin{equation}
\text{d}: \{0,1,-1\} \times \{0,1,-1\} \rightarrow \{0,1\}, \quad  d(a,b)=  \left\{\begin{array}{l}
1, \text{ if } a\neq0\text{ \& }b\neq0\text{ \& }a\neq b\\
0, \text{ otherwise.} \end{array}\right.
\end{equation}
This function is a distance if the following four conditions are satisfied \cite{krey}:

\begin{equation}
\begin{array}{l}
\text{1. }D(\boldsymbol a,\boldsymbol b)\geq 0\\
\text{2. }\boldsymbol a=\boldsymbol b\Leftrightarrow  D(\boldsymbol a,\boldsymbol b)=0\\
\text{3. }D(\boldsymbol a,\boldsymbol b)=D(\boldsymbol b,\boldsymbol a) \text{ (Symmtery)} \\
\text{4. }D(\boldsymbol a,\boldsymbol c) \leq D(\boldsymbol a,\boldsymbol b) +D(\boldsymbol b,\boldsymbol c) \text{ (Triangle inequality)}\\
\end{array}		
\end{equation}
However, we show that this is not always the case and the EHD is an improper distance metric.

\quad
\begin{enumerate}
	\item The first condition is always true due to the definition of EHD which is a summation over a series of $\{0,1\}$ and therefore is always nonnegative.
	\item When $\boldsymbol a=\boldsymbol b$, then for all $j$, $d(\boldsymbol a(j),\boldsymbol b(j))=0$, which leads to the RHS result. However, the reverse is not always true. As an instance, for $\boldsymbol a =[0 1 0]$ and $ \boldsymbol b =[-1 1 0]$, we have $D(\boldsymbol a,\boldsymbol b)=0$, while $\boldsymbol a$ and $\boldsymbol b$ are unequal.
It is concluded that when $D(\boldsymbol a,\boldsymbol b)=0$, for the position of zero entries of $ \boldsymbol a$, the corresponding entries of $ \boldsymbol b$ can be either $1$ or $-1$. Furthermore, this condition forces the corresponding nonzero entries of the two vectors to be equal. Therefore, this condition is true when all the entries are nonzero.
	\item It is obvious that the EHD is symmetric due to the symmetry of $d(\cdot,\cdot)$.
	\item The triangle inequality does not hold. A counter example is:

\begin{equation*}
\boldsymbol a=[1 1 1], \boldsymbol b=[0 1 0], \boldsymbol c=[-1 0 1] \hspace{0.5cm} \Rightarrow  \hspace{0.5cm}
D(\boldsymbol a,\boldsymbol c)=1, D(\boldsymbol a,\boldsymbol b)=0, D(\boldsymbol b,\boldsymbol c)=0,
\end{equation*} 
which yields to an unacceptable result $1\leq 0+0$.
In fact, this condition holds when the locations of zero entries of $\boldsymbol a$, $\boldsymbol b$ and $\boldsymbol c$ are similar or  all the entries are nonzero. Since EHD is part of the MEC function, the provided material in this section gives us an insight to understand the behaviour of MEC.

\end{enumerate}

\section{Proof of Theorem 1}
First, we discuss the fragment matrix model and then show in a lemma that changing the origin of each read does not affect the MEC function. Then a proof of Theorem 1 is given. 

\subsection*{Fragment matrix model}
As introduced in Methods Section, the fragment matrix model is given by 

\begin{equation}
\boldsymbol R=P_\Omega(\boldsymbol M)+\boldsymbol E,
\end{equation}
where $\boldsymbol{E}$ is the error matrix discussed before and $P_\Omega$ is defined in (2). The completed matrix $\boldsymbol M $ is expressed as
\begin{equation}
\boldsymbol M=\boldsymbol u^T \boldsymbol h_{ex}.
\end{equation}
Therefore the rank of matrix $\boldsymbol M$ is one.
Each entry of the $1\times N$ origin vector $\boldsymbol u$ shows the haplotype from which each read originates that can be  either $+1$ or $-1$ corresponding to the  paternal or maternal haplotype, respectively.


\subsection*{Independency of MEC from origin of read}

\noindent Lemma: Consider a given haplotype $\boldsymbol h_{t}$ and two fragment matrices $\boldsymbol R_a$ and $\boldsymbol R_b$ corresponding to two different origin vectors $\boldsymbol u_a$ and $\boldsymbol u_b$. We claim that if the error positions of the fragment matrices are the same, then both matrices are with equal MEC. This is presented in mathematical notation in (B.3).
\begin{equation}
 \forall  \boldsymbol h_t \quad \text{MEC}(\boldsymbol R_a,\boldsymbol h_t) = \text{MEC}(\boldsymbol R_b,\boldsymbol h_{t}),
\end{equation}
where
\begin{eqnarray}
\boldsymbol R_a= P_\Omega(\boldsymbol u_a^T \boldsymbol h)+\boldsymbol E_a,\\
\boldsymbol R_b= P_\Omega(\boldsymbol u_b^T \boldsymbol h)+\boldsymbol E_b,
\end{eqnarray}
in which $\boldsymbol E_a$ and $\boldsymbol  E_b$ are the error matrices whose error positions are identical. 

\quad

\noindent Proof. Since, each row of the fragment matrix  affects the MEC independently, it is enough to prove the lemma only for the $n^{th}$ arbitrary row. To do so, we should prove that
\begin{equation}
\text{min}\{D(\boldsymbol r_a,\boldsymbol h_t),D(\boldsymbol r_a,-\boldsymbol h_t)\}=\text{min}\{D(\boldsymbol r_b,\boldsymbol h_t),D(\boldsymbol r_b,-\boldsymbol h_t)\},
\end{equation}
in which $\boldsymbol r_a$ and  $\boldsymbol r_b$  are the $n^{th}$ rows of $\boldsymbol R_a$ and $\boldsymbol R_b$, respectively defined as
\begin{eqnarray}
\boldsymbol r_a= P_{\Omega_n}(\boldsymbol u_a(n) \boldsymbol h)+\boldsymbol e_a, \\
\boldsymbol r_b= P_{\Omega_n}(\boldsymbol u_b(n) \boldsymbol h)+\boldsymbol e_b,
\end{eqnarray}
where $\boldsymbol u_a(n)$ and $\boldsymbol u_b(n)$ show the $n^{th}$ entries of $\boldsymbol u_a$ and $\boldsymbol u_b$ and $\boldsymbol e_a$  and $\boldsymbol e_b$ are the $n^{th}$ rows of $\boldsymbol E_a$ and $\boldsymbol E_b$, respectively. Also, $P_{\Omega_n}(\cdot)$ is a sub-operator of $P_\Omega(\cdot)$  dedicated to the $n^{th}$  row of a given matrix.
Clearly, for $\boldsymbol u_a(n)= \boldsymbol u_b(n)$, we have $\boldsymbol r_a= \boldsymbol r_b$ and (B.6) is held.  Otherwise, for $\boldsymbol u_a(n) \neq \boldsymbol u_b(n)$, without loss of generality, we assume that $\boldsymbol u_a(n)=1$ and $ \boldsymbol u_b(n)=-1$. Then, (B.8) and (B.9) reduce to
\begin{eqnarray}
\boldsymbol r_a= P_{\Omega_n}(\boldsymbol h )+\boldsymbol e_a \\
\boldsymbol r_b= P_{\Omega_n}(-\boldsymbol h )+\boldsymbol e_b.
\end{eqnarray}

Considering that the  error positions of  $\boldsymbol r_a$ and $\boldsymbol r_b$ are identical and exploiting the model of (B.3), it can be shown that $\boldsymbol e_a=-\boldsymbol e_b$. Using this result in (B.10) and (B.11), we get $\boldsymbol r_a=-\boldsymbol r_b$.
Using the first property of MEC (as shown in Appendix C),  we get
\begin{equation}
D(\boldsymbol r_a,-\boldsymbol h_t)=l_n -D(\boldsymbol r_a,\boldsymbol h_t)=l_n -D(\boldsymbol r_b,- \boldsymbol h_t)=D(\boldsymbol r_b,\boldsymbol h_t),
\end{equation}
where $l_n$ shows the number of nonzero entries of the $n^{th}$  row. By using (B.11) in the left side of (B.6), the lemma is proved as
\begin{equation}
\text{min}\{D(\boldsymbol r_a,\boldsymbol h_t),D(\boldsymbol r_a,-\boldsymbol h_t)\}=\text{min}\{D(\boldsymbol r_a,\boldsymbol h_t),l_n - D(\boldsymbol r_a,\boldsymbol h_t)\}=\text{min}\{D(\boldsymbol r_b,-\boldsymbol h_t),D(\boldsymbol r_b,\boldsymbol h_t)\}.
\end{equation}
This lemma which will be used in the next section shows that the MEC function is not sensitive to the changes of entries of $\boldsymbol u$ under the assumptions of the lemma.

\subsection*{Proof of Theorem 1}
To prove Theorem 1, we propose a specific haplotype vector like $\boldsymbol h_d$ which leads to a lower MEC than the exact haplotype, meaning  that minimizing the MEC function does not necessarily  lead to  the exact haplotype.
 To do so, first we suppose that the antecedent of Theorem 1 is held for the $k^{th}$ column, $i.e.$, $E^{(k)}>c^{(k)}/2$ and $\boldsymbol h_d$ is constructed as
\begin{equation}
\boldsymbol h_d = [ \boldsymbol h_{ex}(1),\ldots,\boldsymbol h_{ex}(k-1), -\boldsymbol h_{ex}(k),\boldsymbol h_{ex}(k+1),\ldots,\boldsymbol h_{ex}(l)],
\end{equation}
in which  $\boldsymbol h_{d}(k)$ is equal to $-\boldsymbol h_{ex}(k)$.
To prove that  $\text{MEC}(\boldsymbol R,\boldsymbol h_d)< \text{MEC}(\boldsymbol R,\boldsymbol h_{ex})$,
  without loss of generality, based on provided lemma, we may consider $\boldsymbol u=[1,\ldots,1]$. In such case, there is no difference to count the mismatches for either the rows or columns of $\boldsymbol R$ due to the definition of MEC in (4). Therefore, we can write MEC as
\begin{equation}
\text{MEC}(\boldsymbol R,\boldsymbol h)= \text{MEC}(\boldsymbol R^{(\sim k)},\boldsymbol h^{(\sim k)})+\text{MEC}(\boldsymbol r^{(k)},\boldsymbol h(k)),
\end{equation}
in which $\boldsymbol r^{(k)}$ is the $k^{th}$ column of $\boldsymbol R$ and $\boldsymbol R^{(\sim k)}$ shows a matrix whose $k^{th}$ column  has been omitted. Furthermore, based on the properties of the MEC function, as shown in Appendix C, it can be seen that
\begin{equation}
\text{MEC}(\boldsymbol r^{(k)},\boldsymbol h(k))=E^{(k)},
 \end{equation}
\begin{equation}
\text{MEC}(\boldsymbol r^{(k)},-\boldsymbol h(k))=c^{(k)}-E^{(k)}.
\end{equation}
Therefore, using (B.15) and (B.16) in (B.14) for $\boldsymbol h_d$  and $\boldsymbol h_{ex}$, we get
\begin{equation}
\text{MEC}(\boldsymbol R,\boldsymbol h_d)= \text{MEC}(\boldsymbol R^{(\sim k)},\boldsymbol h_{ex})+c^{(k)}-E^{(k)},
\end{equation}
\begin{equation}
\text{MEC}(\boldsymbol R,\boldsymbol h_{ex})= \text{MEC}(\boldsymbol R^{(\sim k)},\boldsymbol h_{ex})+E^{(k)}.
\end{equation}
On the other hand, the antecedent of Theorem 1 results in $c^{(k)}-E^{(k)}<E^{(k)}$. Thus, (B.17) and (B.18) accomplish the proof.

\section{Properties of MEC}
The MEC function is calculated for the fragment matrix $\boldsymbol R$ with the dimension of $N \times l$ and the haplotype $\boldsymbol h$ as: 

\begin{equation}
\text{MEC}: \{0,1,-1\}^{N \times l}\times \{1,-1\}^l \rightarrow \mathbb{R}^+\cup \{0\},  \text{MEC}(\boldsymbol R,\boldsymbol  h)= \sum_{i=1}^{N} \text{min}\{D(\boldsymbol r_i,\boldsymbol h),D(\boldsymbol r_i,-\boldsymbol h)\},
\end{equation}
where $D(\cdot,\cdot)$ is defined by (4) and (5) and  $\boldsymbol r_i$ shows the $i^{th}$ row of $\boldsymbol R$.
The following properties can be shown for the MEC function.

\begin{enumerate}
\item For  $\boldsymbol r_i$ with $l_i$ known nonzero entries, by supposing $D(\boldsymbol r_i,\boldsymbol h)=D_i$, we get
\begin{equation}
\begin{array}{l}	
D(\boldsymbol r_i,-\boldsymbol h) =l_i-D_i,\\
D(-\boldsymbol r_i,\boldsymbol h) =l_i-D_i,\\
D(-\boldsymbol r_i,-\boldsymbol h) =D_i.
\end{array}
\end{equation}

\item For every $\boldsymbol r_i$ and  $\boldsymbol h$, we have
\begin{equation}
D(\boldsymbol r_i,\boldsymbol h)= \left\{\begin{array}{l}
k, \text{ if the $i^{th}$ read came from paternal haplotype, }\\
l_i-k, \text{ if the $i^{th}$ read came from maternal haplotype, } \end{array}\right.
\end{equation}
where $k \in N $. For the exact haplotype $\boldsymbol h_{ex}$, $k$ is equal to  the number of error entries  of $\boldsymbol r_i$ denoted by $e_i$. Then, the MEC for the exact haplotype is

\begin{equation}
\text{MEC}(\boldsymbol R,\boldsymbol h_{ex})=\sum_{i=1}^{N} \min\{e_i,l_i-e_i\}=E, 
\end{equation}
in which $E$ is the total number of error in the fragment matrix. 
 For the error-free fragment matrix and the paternal exact haplotype $\boldsymbol h_{ex}$, (C.4) reduces to  $\text{MEC}(\boldsymbol R,\boldsymbol h_{ex})=0$.

\end{enumerate}

\section{Simulation results}

\begin{figure}[!ht]
\center
\begin{subfigure}[b]{0.32\textwidth}
\includegraphics[width=5cm,height=4cm]{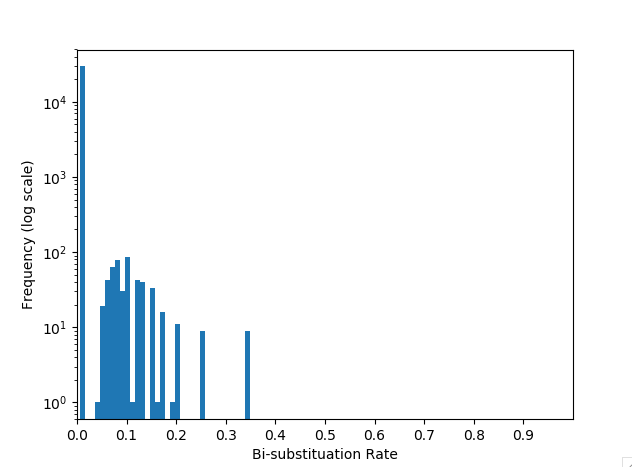} 
\caption{Coverage 10}
\end{subfigure}
\begin{subfigure}[b]{0.32\textwidth}
\includegraphics[width=5cm,height=4cm]{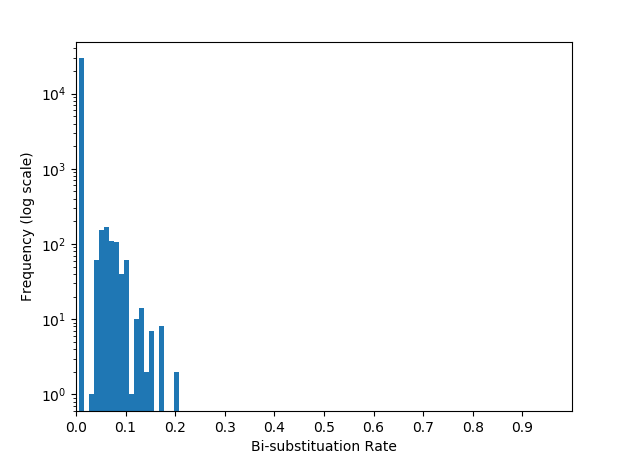}
\caption{Coverage 15 }
\end{subfigure}
\begin{subfigure}[b]{0.32\textwidth}
\includegraphics[width=5cm,height=4cm]{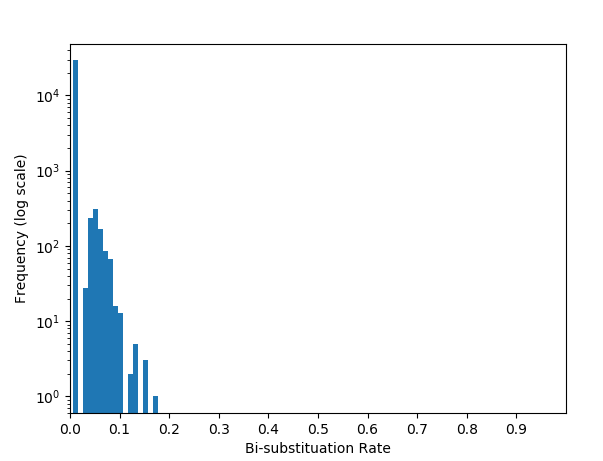}
\caption{Coverage 20}
\end{subfigure}
\caption{Histogram of bi-substitution rates for Illumina reads.}
\end{figure}

\begin{figure}[!ht]
\center
\begin{subfigure}[b]{0.32\textwidth}
\includegraphics[width=5cm,height=4cm]{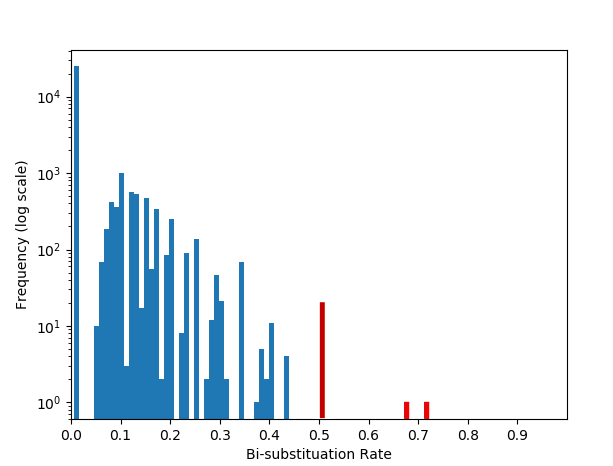}
\caption{Coverage 10}
\end{subfigure}
\begin{subfigure}[b]{0.32\textwidth}
\includegraphics[width=5cm,height=4cm]{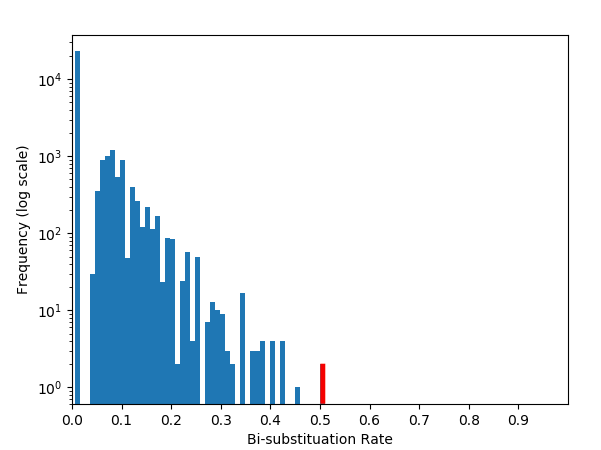}
\caption{Coverage 15}
\end{subfigure}
\begin{subfigure}[b]{0.32\textwidth}
\includegraphics[width=5cm,height=4cm]{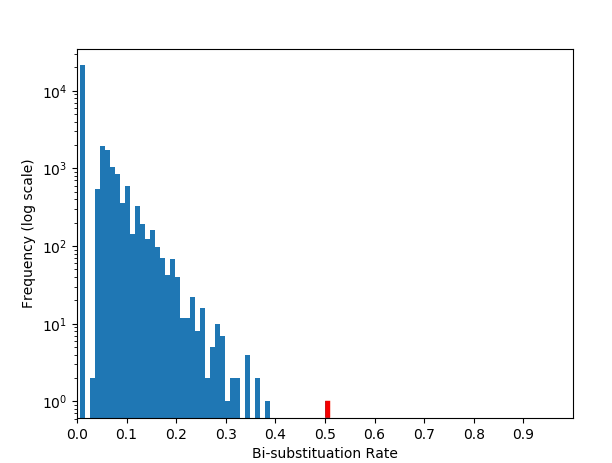}
\caption{Coverage 20 }
\end{subfigure}
\begin{subfigure}[b]{0.32\textwidth}
\includegraphics[width=5cm,height=4cm]{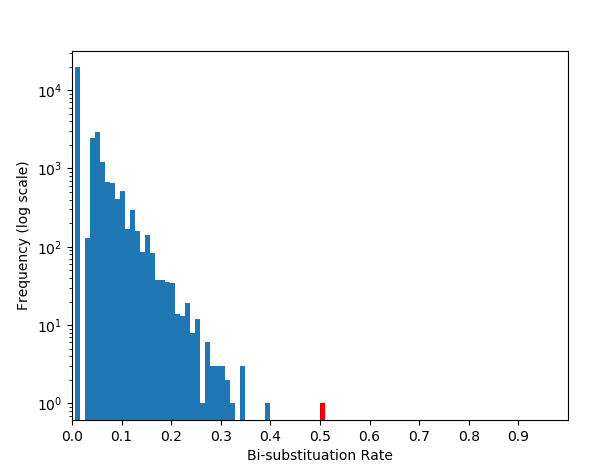}
\caption{Coverage 25 }
\end{subfigure}
\begin{subfigure}[b]{0.32\textwidth}
\includegraphics[width=5cm,height=4cm]{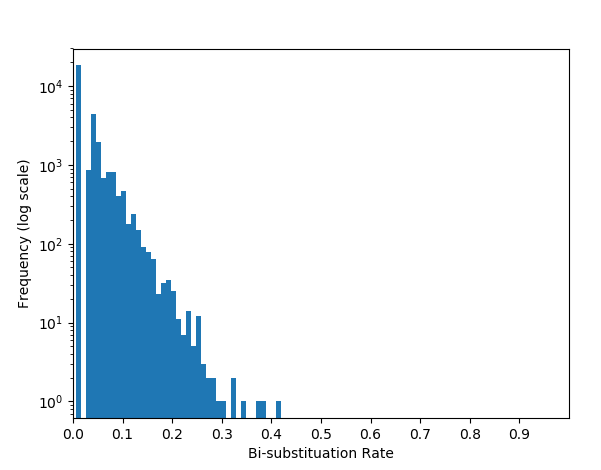}
\caption{Coverage 30 }
\end{subfigure}
\begin{subfigure}[b]{0.32\textwidth}
\includegraphics[width=5cm,height=4cm]{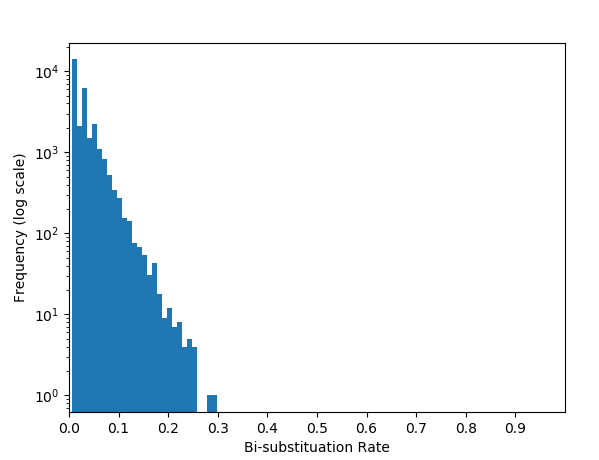}
\caption{Coverage 45 }
\end{subfigure}
\caption{Histogram of bi-substitution rates for PacBio reads. The red bars indicate results for which the antecedent of Theorem 1 is satisfied.}
\end{figure}

\newpage

\section{Polyploid genomes}

Some animals and plants are polyploids. They contain more than two copies of each chromosome. In such case, our modeling presented in (3) and (B.2) may be generalized to 
\begin{equation}
\boldsymbol M=\boldsymbol U^T \boldsymbol H_{ex},
\end{equation}
in which $\boldsymbol H_{ex}$ contains $P$ haplotypes and  $\boldsymbol U$ shows the haplotypic origin of each read. 
The definition of MEC for fragment matrix $\boldsymbol R$ and candidate haplotype $\boldsymbol  H_c$ can be generalized to
\begin{equation}
\text{MEC}(\boldsymbol R,\boldsymbol  H_c)= \sum_{i=1}^{N} \min_{p} D(\boldsymbol r_i,\boldsymbol H_{c}\{p\}),
\end{equation}
in which $\boldsymbol H_{c}\{p\}$ is the $p^{th}$ haplotype (i.e. $p^{th}$ row of  $\boldsymbol H_{c}$).

Here, we consider Theorem 1 (presented in the Methods section) for the polyploid case to show that MEC failure can happen in a specific polyploid case as well. Suppose that the number of reads originating from the $p^{th}$ haplotype covering the $t^{th}$ SNP is $c^{(t)}\{p\}$ and the number of erroneous entries in the $t^{th}$ SNP is $E^{(t)}\{p\}>c^{(t)}\{p\}/2$.

We define a haplotype matrix $\boldsymbol H_d$ and then show that its MEC is lower than that of the exact haplotype $\boldsymbol H_{ex}$.  These two matrices are the same except the $(p,t)^{th}$ element is such that  $\boldsymbol H_{d}(p,t)=-\boldsymbol H_{ex}(p,t)$. In a simplified case, suppose that $P$ haplotypes are separated enough such that the change in one element does not affect the estimation of haplotypic origin of the  reads (the value of $\text{argmin}_{p} D(\boldsymbol r_i,\boldsymbol H_{c}\{p\})$. Thus, due to the definition of MEC in (E.2), we can write

\begin{equation}
\text{MEC}(\boldsymbol R,\boldsymbol  H_d) =\text{MEC}(\boldsymbol R\{\sim p\},\boldsymbol  H_{d}\{\sim p\})+\sum_{i} D(\boldsymbol r_i,\boldsymbol  H_{d}\{p\}),
\end{equation}
in which $\boldsymbol R\{\sim p\}$ is the submatrix of $\boldsymbol R$ restricted to those rows originating from all $P$ haplotypes except $p^{th}$. Last term is a summation over those rows originating from $p^{th}$ haplotype. Due to the definition of $\boldsymbol  H_d$ and $\boldsymbol  H_{ex}$, we have 
\begin{equation}
\text{MEC}(\boldsymbol R\{\sim p\},\boldsymbol  H_{d}\{\sim p\})=\text{MEC}(\boldsymbol R\{\sim p\},\boldsymbol  H_{ex}\{\sim p\}).
\end{equation}

Due to the assumption on erroneous entries, we have $\sum_{i} D(\boldsymbol r_i,\boldsymbol  H_{d}\{p\}) < \sum_{i} D(\boldsymbol r_i,\boldsymbol  H_{ex}\{p\})$ which results in 

\begin{equation}
\text{MEC}(\boldsymbol R,\boldsymbol  H_{d}) < \text{MEC}(\boldsymbol R,\boldsymbol  H_{ex}).
\end{equation}
This shows that optimizing the MEC function does not guarantee reaching the exact (true) haplotypes.

\quad

\bibliographystyle{unsrt}

\begin{thebibliography}{1}

\bibitem{sny}
Snyder MW, Adey A, Kitzman JW, Shendure J.
\newblock Haplotype-resolved genome sequencing: experimental methods and  applications.
\newblock {\em Nature Reviews Genetics}, 16(6):344, 2015.

\bibitem{schwartz}
Schwartz R.
\newblock Theory and algorithms for the haplotype assembly problem.
\newblock {\em Communications in Information \& Systems}, 10(1):23--38, 2010.

\bibitem{shen} 
Shendure J, Balasubramanian S, Church GM, Gilbert W, Rogers J, Schloss JA, et al.
\newblock DNA sequencing at 40: past, present and future.
\newblock {\em Nature}, 550(7676):345, 2017.

\bibitem{goodwin}
Goodwin S, McPherson JD, and McCombie WR.
\newblock Coming of age: ten years of next-generation sequencing technologies.
\newblock {\em Nature Reviews Genetics}, 17(6):333--351, 2016.


\bibitem{haptree}
Berger E, Yorukoglu D, Peng J, Berger B.
\newblock HapTree: A novel Bayesian framework for single individual polyplotyping using NGS data.
\newblock {\em PLoS Computational Biology}, 10(3):e1003502, 2014.

\bibitem{bansal}
Bansal V, Halpern A, Axelrod N, Bafna V.
\newblock An MCMC algorithm for haplotype assembly from whole-genome sequence data.
\newblock {\em Genome Research}, 18(8):1336--1346, 2008.


\bibitem{lancia}
Lancia G, Bafna V, Istrail S, Lippert R, Schwartz R.
\newblock SNPs problems, complexity, and algorithms.
\newblock  ESA 2001. Lecture Notes in Computer Science, vol 2161. Springer, Berlin, Heidelberg. 182--193. 2001.


\bibitem{hapcut}
Bansal V, Bafna V.
\newblock HapCut: an efficient and accurate algorithm for the haplotype assembly problem.
\newblock {\em Bioinformatics}, 24(16):i153--i159, 2008.

\bibitem{wang05}
Wang RS, Wu LY, Li ZP, Zhang XS.
\newblock Haplotype reconstruction from SNP fragments by minimum error correction.
\newblock {\em Bioinformatics}, 21(10):2456--2462, 2005.

\bibitem{chen}
Chen ZZ, Deng F, Wang L.
\newblock Exact algorithms for haplotype assembly from whole-genome sequence data.
\newblock {\em Bioinformatics}, 29(16):1938--1945, 2013.

\bibitem{das}
Das  S, Vikalo H.
\newblock SDHaP: haplotype assembly for diploids and polyploids via semi-definite programming.
\newblock {\em BMC Genomics}, 16(1):260, 2015.

\bibitem{kuleshov}
Kuleshov V.
\newblock Probabilistic single-individual haplotyping.
\newblock {\em Bioinformatics}, 30(17):i379--i385, 2014.

\bibitem{he}
He D, Choi A, Pipatsrisawat K, Darwiche A, Eskin E.
\newblock Optimal algorithms for haplotype assembly from whole-genome sequence data.
\newblock {\em Bioinformatics}, 26(12):i183--i190, 2010.

\bibitem{deng}
Deng F, Cui W, Wang L.
\newblock A highly accurate heuristic algorithm for the haplotype assembly problem.
\newblock {\em BMC genomics}, 14(2):S2, 2013.

\bibitem{bonizzoni}
Bonizzoni P, Dondi R, Klau GW, Pirola Y, Pisanti N, Zaccaria S. 
\newblock On the minimum error correction problem for haplotype assembly in  diploid and polyploid genomes.
\newblock {\em Journal of Computational Biology}, 23(9):718--736, 2016.

\bibitem{zhang}
Zhang XS, Wang RS, Wu LY, Zhang W. 
\newblock Minimum conflict individual haplotyping from SNP fragments and related genotype.
\newblock {\em Evolutionary Bioinformatics Online}, 2:261, 2006.

\bibitem{refhap}
Duitama J, McEwen GK, Huebsch T, Palczewski S, Schulz S, Verstrepen K, et al.
\newblock Fosmid-based whole genome haplotyping of a hapmap trio child:  evaluation of single individual haplotyping techniques.
\newblock {\em Nucleic Acids Research}, 40(5):2041--2053, 2011.

\bibitem{si}
Si H, Vikalo H, Vishwanath S. 
\newblock Information-theoretic analysis of haplotype assembly.
\newblock {\em IEEE Transactions on Information Theory}, 63(6):3468--3479,
  2017.

\bibitem{maj}
Majidian S, Kahaei MH.
\newblock NGS based haplotype assembly using matrix completion.
\newblock {\em PLoS ONE}, 14(3): e0214455, 2019.

\bibitem{lander}
Lander ES, Waterman MS.
\newblock Genomic mapping by fingerprinting random clones: a mathematical analysis.
\newblock {\em Genomics}, 2(3):231--239, 1988.

\bibitem{klambauer}
Klambauer G, Schwarzbauer K, Mayr A, Clevert DA, Mitterecker A, Bodenhofer U,  et al.
\newblock cn.MOPS: mixture of poissons for discovering copy number variations in next-generation sequencing data with a low false discovery rate.
\newblock {\em Nucleic Acids Research}, 40(9):e69--e69, 2012.

\bibitem{edg}
Edge P, Bansal V.
\newblock Longshot: accurate variant calling in diploid genomes using single-molecule long read sequencing.
\newblock {\em bioRxiv}, 1:564443, 2019.


\bibitem{tyl}
Tyler AD, Mataseje L, Urfano CJ, Schmidt L, Antonation KS, Mulvey MR, et al.
\newblock  Evaluation of Oxford Nanopore’s MinION Sequencing Device for Microbial Whole Genome Sequencing Applications.
\newblock {\em Scientific Reports}, 8(1):10931, 2018.

\bibitem{levy}
Levy S, Sutton G, Ng PC, Feuk L, Halpern AL, Walenz BP, et al.
\newblock The diploid genome sequence of an individual human.
\newblock {\em PLoS Biology}, 5(10):e254, 2007.

\bibitem{grch}
Schneider VA, Graves-Lindsay T, Howe K, Bouk N, Chen HC, Kitts PA, et al.
\newblock Evaluation of GRCh38 and de novo haploid genome assemblies demonstrates the enduring quality of the reference assembly.
\newblock {\em Genome Research}, 27(5):849--64, 2017.

\bibitem{mot}
Motazedi E, Finkers R, Maliepaard C, de Ridder D.
\newblock Exploiting next-generation sequencing to solve the haplotyping puzzle in polyploids: a simulation study.
\newblock {\em Briefings in Bioinformatics}, 19(3):387--403, 2017.

\bibitem{pb}
Ono Y, Asai K, Hamada M.
\newblock  PBSIM: PacBio reads simulator - toward accurate genome assembly.
\newblock {\em Bioinformatics}, 29(1):119--21, 2012.

\bibitem{mini}
Li H.
\newblock  Minimap2: pairwise alignment for nucleotide sequence
\newblock {\em Bioinformatics}, 4(18):3094--100, 2018.

\bibitem{art}
Huang W, Li L, Myers JR, Marth GT.
\newblock  ART: a next-generation sequencing read simulator.
\newblock {\em Bioinformatics}, 28(4):593--4, 2011.

\bibitem{bwa}
Li H, Durbin R.
\newblock  Fast and accurate short read alignment with Burrows–Wheeler transform.
\newblock {\em Bioinformatics}, 25(14):1754--60, 2009.

\bibitem{freeb}
Garrison, E., and Marth, G.
\newblock Haplotype-based variant detection from short-read sequencing.
\newblock {\em arXiv preprint},.  1207.3907,  2012.

\bibitem{sam}
Li H, Handsaker B, Wysoker A, Fennell T, Ruan J, Homer N, Marth G, Abecasis G, Durbin R.
\newblock The sequence alignment/map format and SAMtools.
\newblock {\em Bioinformatics}, 25(16):2078--9, 2009.

\bibitem{trip}
Motazedi E, de Ridder D, Finkers R, Baldwin S, Thomson S, Monaghan K, Maliepaard C. 
\newblock  TriPoly: haplotype estimation for polyploids using sequencing data of related individuals.
\newblock {\em Bioinformatics},. 34(22):3864--72, 2018.

\bibitem{por}
Porubsky D, Garg S, Sanders AD, Korbel JO, Guryev V, Lansdorp PM, Marschall T.
\newblock Dense and accurate whole-chromosome haplotyping of individual genomes.
\newblock {\em Nature Communications}, 8(1):1293, 2017.

\bibitem{krey}
Kreyszig E.
\newblock {\em Introductory functional analysis with applications}, volume~1.
\newblock Wiley, 1989.

\end{thebibliography}

\end{document}